\documentclass[photonics,article,submit,moreauthors,pdftex,10pt,a4paper]{mdpi} 

\firstpage{1} 
\articlenumber{x}
\doinum{10.3390/------}
\pubvolume{xx}
\pubyear{2016}
\copyrightyear{2016}
\externaleditor{Academic Editors: Jens Biegert and Peter G\"unter}
\history{Received: date; Accepted: date; Published: date}

\pdfoutput=1

\Title{Constant Matrix Element Approximation to Time-Resolved Angle-Resolved Photoemission Spectroscopy}

\Author{J. K. Freericks$^{1,\ddagger}$* and H. R. Krishnamuthy$^{2,3,\ddagger}$}
\AuthorNames{J. K. Freericks and H. R. Krishnamurthy}

\address{%
$^{1}$ \quad Department of Physics, Georgetown University, 37$^{\rm th}$ and O Sts. NW, Washington, DC 20057, U.S.A.; james.freericks@georgetown.edu\\
$^{2}$ \quad Centre for Condensed Matter Theory, Department of Physics, Indian Institute of Science,
Bangalore 560012, India ; hrkrish@gmail.com\\
$^3$ \quad Condensed Matter Theory Unit, Jawaharlal Nehru Centre for Advanced Scientific Research,
Bangalore 560064, India}

\corres{Correspondence: james.freericks@georgetown.edu; Tel.: +202-687-6159}

\secondnote{These authors contributed equally to this work.}

\abstract{We describe different approximations associated with employing a constant matrix element for the coupling of light to multiband electrons in the context of time-resolved angle-resolved photoemission spectroscopy (TR-ARPES). In particular, we demonstrate that the constant matrix approximation only holds for one of the coordinate representations, and changing to other bases, requires including 
nonconstant corrections to the matrix element. We also discuss some simplifying approximations, where a constant matrix element is employed in multiple bases, and the consequences of this further approximation (especially with respect to the TR-ARPES signal no longer being nonnegative). We also discuss issues related to gauge invariance of the final spectra.}


\begin{document}



\section{Introduction}

Time-resolved and angle-resolved photoemission spectroscopy (TR-ARPES) is emerging as one of the powerful tools in ultrafast 
optics to probe the nonquilibrium behavior of quantum materials and how they recover back to a quasiequilibrium state. 
Many different materials have been investigated so far, including charge-density-wave systems~\cite{perfetti1,perfetti2,rossnagel,rossnagel2,schmitt,schmitt2,bovensiepen,mihailovich1,mihailovich2}, high-temperature superconductors~\cite{lanzara1,lanzara2,lanzara3,shen,rameau}, iron pnictides~\cite{bovensiepen2}, and Mott insulators~\cite{wolf}. Theory has also been well developed to model these systems.
The original theory, described the most general formulas for this pump/probe spectroscopy~\cite{fkp_2009} (including corrections to
ensure gauge invariance and including the dominant second-order perturbation theory contributions~\cite{scripta}). Because the
formula for TR-ARPES (in a gauge) comes from second-order perturbation theory, it involves the square of a many-body matrix element, and can be shown to be explicitly nonnegative. The true physical response, however, is the gauge-invariant response, and that can be constructed following a straightforward procedure~\cite{bertoncini_jauho}, but it has proved to be difficult to establish, in general, that this result is nonegative~\cite{scripta}. 

It is common to also invoke a constant matrix element approximation within theoretical calculations of photoemission spectra. This is not necessarily because the matrix elements are known to be constant, but is because explicitly determining the matrix elements is extremely challenging for quantum materials, so employing a constant matrix element is a good first step in modeling such systems. The constant matrix element approximation is trivial when there is only a single effective band in the system, but it becomes more complex when there are multiple bands. In this paper, we discuss different options one has for invoking the constant matrix element approximation within TR-ARPES, paying particular attention to the issues of gauge invariance and nonnegativity of the spectra. We also discuss implications for time-resolved photoemission spectroscopy (TR-PES), where the momentum dependence has been integrated over.

\section{Results}

The description of a pump/probe experiment is complicated by the fact that the presence of the pump removes time-translation-invariance symmetry from the system, and we find, generically, that the Green's function depends on two times. Using the Wigner coordinates~\cite{wigner}, these two times can be summarized as the average time $t_{\rm ave}=(t_1+t_2)/2$ and the relative time $t_{\rm rel}=t_1-t_2$. The pump must be included explicitly into the Hamiltonian, because it is a large ``perturbation'' of the system, so it cannot be treated with low-order perturbation theory. Fortunately, because the Peierls substitution~\cite{peierls} allows us to treat the electric-field effects exactly for single-band models, we have an exact means to include the effects of the pump pulse. Note, that in multiband models, there can be additional electric field effects due to symmetry allowed dipole transitions that connect two bands that have opposite spatial parity (and, of course, higher-order multipoles can also be taken into account, but it is rarely needed to do so). The dipole approximation would involve a direct coupling (hybridization) between the two bands, with the strength of the coupling proportional to the field amplitude. This can also be easily included into the Hamiltonian, which then fully describes the effects of the electric field (however, magnetic field effects, including fields generated by time-varying electric fields, are usually neglected). The other constraint is that the field is uniform in space. Nonuniform fields create many additional complications to the formalism. The probe pulse, on the other hand, is usually taken to be weak, and hence it can be treated with lowest order (second-order) perturbation theory. While there are a number of different second-order terms that can contribute, only one term dominates, and that is the only one we will focus on here~\cite{fkp_2009,scripta}.

\subsection{Formalism for the TR-ARPES response in a gauge}

It was demonstrated in Ref.~\cite{fkp_2009}, that the TR-ARPES signal can be found from a probe-pulse-weighted two-time Fourier transform of the nonequilibrium lesser Green's function. We need to review that derivation to first correct it for a proper gauge-invariant treatment, and to establish our notation so we can discuss the constant matrix-element approximation.

We assume the Hamiltonian $\mathcal{H}(t)$ includes all of the interactions of the electrons (with themselves, and with other scatterers, like impurities, phonons, etc.) and it also includes the effect of the electric field, as described above, which makes it time-dependent. The explicit form of this Hamiltonian can be arbitrary for this work, so we do not specify a particular Hamiltonian at this time. The evolution operator of the Hamiltonian with the pump field is denoted $U(t,t_0)$ for the evolution from time $t_0$ to time $t$, and it satisfies the equation of motion $i\hbar d U(t,t_0)/dt=\mathcal{H}(t)U(t,t_0)$; it is a unitary operator that also satisfies the semigroup properties $U(t,t)=1$ and $U(t,t_1)U(t_1,t_0)=U(t,t_0)$. The photocurrent operator ${\bf J}_d$ represents the current of electrons at the spatial location of the detector (given by ${\bf R}_d$) and with momentum peaked around ${\bf k}_e$. We let $\phi_{{\bf k}_e;{\bf R}_d}({\bf k})$ be a momentum-space wavefunction peaked about both ${\bf k}_e$ and ${\bf R}_d$ and we let $c^\dagger_{{\bf k}_e;{\bf R}_d}$ be the associated fermionic creation operator to create fermions in the state $\phi$. Then the photocurrent operator satisfies ${\bf J}_d\approx (\hbar {\bf k}_e/m_e)c^\dagger_{{\bf k}_e;{\bf R}_d}c^{\phantom{\dagger}}_{{\bf k}_e;{\bf R}_d}$, with $m_e$ the mass of the electron. Note that we are suppressing the spin degree of freedom for notational simplicity.

Under the assumption that $\mathcal{H}_{\rm probe}(t)$ is a weak perturbation to probe the nonequilibrium system, the dominant second-order contribution to the photocurrent becomes
\begin{equation}
\langle {\bf J}_d(t)\rangle=\frac{1}{\hbar^2}\int_{t_0}^tdt_1\int_{t_0}^tdt_2\langle U(-\infty,t_2)\mathcal{H}_{\rm probe}(t_2)U(t_2,t){\bf J}_dU(t,t_1)\mathcal{H}_{\rm probe}(t_1)
U(t_1,-\infty)\rangle,
\end{equation}
with
\begin{equation}
\langle \mathcal{O}\rangle=\frac{1}{\mathcal{Z}}\sum_n e^{-\beta E_n}\langle n|\mathcal{O}|n\rangle,
\end{equation}
and $\mathcal{Z}=\sum_n\exp(-\beta E_n)$. Here, $E_n$ is the energy of a many-body eigenstate $|n\rangle$ of the
Hamiltonian $\mathcal{H}(-\infty)$ before the pump is turned on,  $\beta=1/k_BT$ is the inverse temperature 
of the initial equilibrium configuration prior to the pump being turned on, and $\mathcal{O}$ is any operator.

Our material is described as a bulk system with a surface parallel to the $x$-$y$ plane. We let $\epsilon_{\nu{\bf k}_\parallel}$ denote the bandstructure with momentum ${\bf k}_\parallel$ parallel to the surface, and $\nu$ denotes all the other quantum numbers (like band index, etc.). The bandstructure is calculated for the initial Hamiltonian $\mathcal{H}(-\infty)$ prior to the pump being turned on. We let $c^\dagger_{\nu{\bf k}_\parallel}$ denote the corresponding electron creation operator; note that we are continuing to suppress spin. The component of $\mathcal{H}(t)$ that corresponds to the absorption of a photon with wavevector ${\bf q}$, frequency $\omega_{\bf q}=cq$, and creation operator $a^\dagger_{\bf q}$, becomes
\begin{equation}
\sum_{\nu\nu'{\bf k}_\parallel}s(t)e^{-i\omega_{\bf q}t}M_{\bf q}(\nu,\nu',{\bf k}_\parallel; t)c^\dagger_{\nu'{\bf k}_\parallel+{\bf q}_\parallel} c^{\phantom{\dagger}}_{\nu{\bf k}_\parallel}a^{\phantom{\dagger}}_{\bf q}
\end{equation}
with $M_{\bf q}(\nu,\nu',{\bf k}_\parallel; t)=\langle \nu^\prime \mathbf{k}^\prime_\parallel  |(i e \hbar {\mathbf{A}}_{probe}(\mathbf{r})/m_e c)\cdot (\nabla - i e  {\mathbf{A}}_{pump}(\mathbf{r}, t)/(\hbar c)) | \nu \mathbf{k}_\parallel \rangle$ the one-electron matrix element. Here, ${\bf A}_{probe}({\bf r})$ is a product of a spatial envelope function and a plane wave of wavevector ${\bf q}$ and $s(t)$ is the probe pulse temporal envelope function. The matrix element itself
depends on the vector potential of the pump field and hence inherits its time dependence. If a tight binding model with a Peierls substitution for the coupling of the system to the vector potential of the pump field is used, the expression for the matrix element $M$ has to be appropriately rewritten.

Introducing the time-reversed LEED (TRL) states, as described in Ref.~\cite{fkp_2009}, we then find the total photocurrent for a particular
probe pulse is given by
\begin{equation}
\langle {\bf J}_d\rangle=\frac{\hbar{\bf k}_e}{m_e}\sum_{\nu\nu'{\bf k}_\parallel{\bf k'}_\parallel}\phi^*_{{\bf k}_e;{\bf R}_d}[{\bf k}(\nu)]\phi_{{\bf k}_e;{\bf R}_d}[{\bf k}'(\nu')]P_{\bf k}(s),
\end{equation}
where $P_{\bf k}(s)$ is the total probability for a photoelectron to be detected at the detector for a given probe pulse envelope function $s(t)$. We find
\begin{eqnarray}
P_{\bf k}(s)&=&\frac{1}{\hbar^2}\sum_{\nu_1\nu_1'{\bf k}_{\parallel 1}}\sum_{\nu_2\nu_2'{\bf k}_{\parallel 2}}
\int_{-\infty}^\infty dt_1\int_{-\infty}^\infty dt_2M^*_{\bf q}(\nu_2,\nu_2',{\bf k}_{\parallel 2};t_2)M_{\bf q}(\nu_1,\nu_1',{\bf k}_{\parallel 1};t_1)s(t_2)s(t_1)e^{i\omega_{\bf q}(t_2-t_1)}\nonumber\\
&\times&\langle c_{\nu_2{\bf k}_{\parallel 2}}^\dagger(t_2)c^{\phantom{\dagger}}_{\nu_2'{\bf k}_{\parallel 2}+{\bf q}_{\parallel}}(t_2)
c_{\nu'{\bf k}_{\parallel}'}^\dagger(t)c^{\phantom{\dagger}}_{\nu{\bf k}_{\parallel}}(t)
c_{\nu_1'{\bf k}_{\parallel 1}+{\bf q}_{\parallel}}^\dagger(t_1)c^{\phantom{\dagger}}_{\nu_1{\bf k}_{\parallel 1}}(t_1)
\rangle.
\end{eqnarray}
All fermionic creation and annihilation operators are in the Heisenberg representation with respect to the Hamiltonian
with the pump present, {\it i.~e.~}$\mathcal{H}(t)$.

The expectation value is a three-particle nonequilibrium many-body Green's function, which is challenging to evaluate. But,
it is a good approximation to assume that the high energy electrons (for example, the TRL state electrons indexed by $\nu$ and $\nu'$) do not strongly interact with other electrons, and that the pump pulse does not significantly alter them (because they reside primarily outside the
region where the pump pulse lives). In this case, they can be treated as noninteracting band-like electrons, which allows the 
three-particle Green's function to be factorized, just like in Wick's theorem, which yields
\begin{eqnarray}
&~&\langle c^{\dagger}_{\nu_2 \mathbf{k}_{\parallel 2}}(t_2) c_{\nu_1 \mathbf{k}_{\parallel 1}}(t_1)\rangle_\mathcal{H} \; \times \; \langle c_{\nu_2^\prime \mathbf{k}_{\parallel 2}+\mathbf{q}_\parallel}(t_2) c^{\dagger}_{\nu^\prime \mathbf{k}_\parallel^\prime}(t)\rangle_{\mathcal{H}_0} \times 
\langle c_{\nu \mathbf{k}_\parallel}(t)  c^{\dagger}_{\nu_1^\prime \mathbf{k}_{\parallel 1}+\mathbf{q}_\parallel}(t_1)\rangle_{\mathcal{H}_0} \nonumber\\
&\simeq &\langle c^{\dagger}_{\nu_2 \mathbf{k}_{\parallel 2}}(t_2) c_{\nu_1 \mathbf{k}_{\parallel 1}}(t_1)\rangle_\mathcal{H} \, \delta_{\nu_2^\prime, \nu^\prime} 
 \delta_{\nu_1^\prime, \nu} \delta_{\mathbf{k}_{\parallel 2}+\mathbf{q}_\parallel, \mathbf{k}_\parallel^\prime}  \delta_{\mathbf{k}_{\parallel 1}+\mathbf{q}_\parallel, \mathbf{k}_\parallel}
e^{i [(\epsilon_{\nu^\prime \mathbf{k}_\parallel^\prime}-\mu)(t-t_2)- (\epsilon_{\nu \mathbf{k}_\parallel} - \mu)(t-t_1)]}
\end{eqnarray}
Here, the expectation value involving the low-energy electrons is evaluated as a single-particle nonequilibrium lesser Green's function evolving according to the full Hamiltonian $\mathcal{H}$. The high-energy averages are noninteracting equilibrium single-particle Green's functions evaluated with respect to the band Hamiltonian $\mathcal{H}_0$, which includes no pump and no interactions of the electrons (with themselves or anything else). These high energy averages, can then be explicitly evaluated, as shown in the last line above, where we have further assumed that the initial temperature is small compared to the energies of the photo-emitted electrons And the pump does not excite the electrons to too high an energy.

A further approximation is to assume that the momentum distribution of the electrons arriving at the detector is sharp. Then we have ${\bf k}_e={\bf k}(\nu)={\bf k}'(\nu')$ and $\nu=\nu'=\nu_e$. Finally, we also have $\epsilon_{\nu_e{\bf k}_{\parallel e}}-\mu=\hbar\omega_{\bf q}-\hbar\omega$, with $\hbar\omega$ the 
excitation energy remaining in the system after the photoemission process. This then yields the main result for the TR-ARPES formalism
\begin{eqnarray}
P_{\bf k}(s)&\simeq& -\frac{i}{\hbar^2}\sum_{\nu_1,\nu_2}\int_{-\infty}^\infty dt_1\int_{-\infty}^\infty dt_2M_{{\bf q}=0}^*(\nu_2,\nu_e;{\bf k}_{e\parallel };t_2)
M_{{\bf q}=0}(\nu_1,\nu_e;{\bf k}_{e\parallel };t_1) s(t_2)s(t_1)e^{i\omega(t_2-t_1)}\nonumber\\
&\times&G^<_{\nu_1{\bf k}_{e\parallel},\nu_2{\bf k}_{e\parallel}}(t_1,t_2),
\label{eq: trarpes}
\end{eqnarray}
where we assumed the photon wavevector is much smaller than the electron wavevectors. The two-time nonequilibrium lesser Green's function is
\begin{equation}
G^<_{\nu_1{\bf k}_{e\parallel},\nu_2{\bf k}_{e\parallel}}(t_1,t_2)=i\langle c^\dagger_{\nu_2{\bf k}_{e\parallel}}(t_2)
c^{\phantom{\dagger}}_{\nu_1{\bf k}_{e\parallel}}(t_1)\rangle_{\mathcal{H}}
\end{equation}
and the operator averages are taken with respect to the full Hamiltonian. We label the photoresponse with the symbol
$s$, since the envelope function of the probe Hamiltonian encodes the time delay with respect to the pulse. Since the integrals over time extend from $-\infty$ to $\infty$, we count {\it all} photoelectrons which reach the detector from the applied pulses.

\subsection{Gauge invariance of the TR-ARPES signal}

In the previous subsection, we calculated the TR-ARPES signal in a specific gauge (the so-called Hamiltonian gauge where
the vector potential is nonzero, but the scalar potential vanishes). If the time delay between the pump and the probe is long enough, that the pump and the probe pulses do not overlap in time, then the vector potential vanishes, and the results are manifestly gauge invariant. In addition,  if one calculates the {\it  total} photoemission response, integrating over momentum, the final result is also manifestly gauge invariant even when the pump and probe pulses do overlap.  However, in this latter case, if one is interested in the {\it angle-resolved} response, then, while the full expression in Eq.~(\ref{eq: trarpes}) is easily shown to be gauge-invariant, one must exercise care with the approximations one inevitably makes, especially the approximation of ignoring the momentum dependence of the matrix elements (which we will refer to as the constant matrix element appoximation), in evaluating the response function to ensure that it continues to be gauge invariant. To date, no one has properly formulated such gauge-invariant prescriptions when effects of a surface,  of a three dimensional dispersing band structure, 
or of the pump pulse on the propagation of the photoelectrons, are important. We won't solve this problem here either. Instead, we focus on the simplest scenario, where (in addition to the constant matrix element approximation) the spatial dependence of the electric field is neglected
(which is an excellent approximation for optical or infrared pump pulses) and where the Green's function depends only on the momenta parallel to the sample surface.

In this case, one possible prescription is to follow Bertoncini and Jauho~\cite{bertoncini_jauho}, and replace the lesser Green's function by its gauge invariant modification:
\begin{equation}
G^<_{\nu_1 \mathbf{k}_{e\parallel}, \nu_2 \mathbf{k}_{e\parallel}}(t_1, t_2) \rightarrow {\tilde{G}}^<_{\nu_1 \mathbf{k}_{e\parallel}, \nu_2 \mathbf{k}_{e\parallel}}(t_1, t_2)  \equiv  G^<_{\nu_1 {\bar{\mathbf{k}}}_{e\parallel}, \nu_2 {\bar{\mathbf{k}}}_{e\parallel}}(t_1, t_2),
\end{equation}
with
\begin{equation}
{\bar{\mathbf{k}}}_{e\parallel}(t_1,t_2) = {\mathbf{k}}_{e\parallel} - \frac{e}{\hbar c} \frac{1}{t_1 - t_2} \int_{-(t_1 - t_2)/2}^{(t_1 - t_2)/2} dt^\prime
{\mathbf{A}}_{pump}(\frac{t_1+t_2}{2}+t^\prime) .
\end{equation}
While it is straightforward to show that the TR-ARPES signal is nonnegative in a gauge, it remains challenging to show nonnegativity for the gauge-invariant signal, although physically, this result must always be true.

\subsection{Constant matrix-element approximation}

For simplicity, we focus on quasi-2d materials, where ${\bf k}_\parallel={\bf k}$ is the 2d wavevector. This approximation works well for many interesting systems including high $T_c$ superconductors, graphene, transition-metal dicalchogenides,
and many topological insulators. If only one band ($\nu_1$) contributes at low energy, then the constant matrix element approximation becomes trivial. Namely, the gauge-invariant TR-ARPES signal becomes
\begin{equation}
P^{GI}_{\bf k}(s)\simeq -\frac{i}{\hbar^2}|M|^2\int_{-\infty}^\infty dt_1\int_{-\infty}^\infty dt_2 s(t_2)s(t_1)e^{i\omega(t_2-t_1)}G^<_{\nu_1\bar{\bf k}_{e}(t_1,t_2),\nu_1\bar{\bf k}_{e}(t_1,t_2)}(t_1,t_2),
\end{equation}
and there is no freedom in the approximation (we dropped the parallel symbol since the wavevector is in 2d only here). As mentioned before, the result in a gauge, is nonnegative, as a trivial application of Bochner's theorem~\cite{bochner,scripta}, but for the gauge-invariant result is difficult to prove nonnegativity.

For higher numbers of bands, the matrix elements, even when assumed to be independent of momentum, can have  different values for different bands. By examining the structure of the photoemission response, we see that it involves a product of a vector (of the different matrix elements) times the lesser Green's function matrix times the Hermitian conjugate of the vector of matrix elements, which is a quadratic form. Such an object is nonnegative if all eigenvalues of the matrix are nonnegative. But, more important than that, is the fact that the result requires a summation over {\it all matrix elements of $G^<$}. Hence, it is not a matrix invariant purely of the $G^<$ matrix; and if one changes the basis, especially if one changes the basis as a function of time, then the matrix elements must also be changed according to the unitary transformation that affects the change of basis, and hence will become functions of time. This is true even if we choose the matrix elements to be the same for all the bands.
Of course, the total photoemission signal is independent of whatever basis is chosen to represent the Green's function matrix. The vector of matrix elements changes its values as one changes the basis so that the final result of the vector-matrix-vector product does indeed remain invariant of the basis. {\it But one cannot assume that the individual matrix elements are constants independent of the basis chosen for the bands.} They can change as the basis changes, and indeed must change for the final result to be invariant.
Emphasizing this subtle point is the main message of this work. 

These results imply that the constant matrix element approximation becomes more complicated when multiple bands are involved in the low-energy dynamics of the material. Before we discuss the implications of these choices further, we first want to discuss the situation if we consider the angle integrated photoemission spectroscopy. In the constant matrix element approximation, we neglect the ${\bf k}_e$ dependence of the matrix element, so the TR-PES signal becomes
\begin{equation}
P^{GI}_{\rm local}(s)\simeq -\frac{i}{\hbar^2}\int_{-\infty}^\infty dt_1\int_{-\infty}^\infty dt_2 s(t_2)s(t_1)e^{i\omega(t_2-t_1)}\sum_{\nu_1\nu_2}M_{\nu_1}G^{<\,{\rm local}}_{\nu_1,\nu_2}(t_1,t_2)M^*_{\nu_2}\,.
\label{eq: final_pes}
\end{equation}
It is common to perform the calculations in a basis that is diagonal in band space. In this case, if we also choose all of the matrix elements to be the same, then the TR-PES signal is given by the trace of the Green's function, which can give the false impression that the result is always the trace of the Green's function matrix. But, as we discussed above, this is not the case.

Sometimes it might be convenient to work in the instantaneous diagonal basis for a given wavevector. Because the Peierls substitution shifts the wavevector in a time-dependent fashion, the basis changes with time. (For example, the Bloch wave functions of the instantaneous one electron component of $\mathcal{H}(t)$ would have this character.)  In this circumstance, one can choose to make a further approximation of choosing the matrix elements to be constant (and thereby factor out of the calculation) even though the basis is being changed for each time point, and use the trace formula, since the Green's function is diagonal in this basis. It should be noted that doing so introduces an uncontrolled approximation into the system and could result in a photoemission spectra which became negative. Nevertheless, it may be an interesting approximation to make if no further information is known about the matrix elements~\cite{graphene,fregoso}.

The general conclusion that we make for the constant matrix element approximation is that it can become problematic when it is applied to systems with more than one band. The best way to proceed, if feasible, is to determine a reasonable guess for the ratios of the matrix elements in the different bands, keep track of all basis changing operations during the calculation, transform the matrix elements accordingly, and be sure to evaluate the final result as the product of the Green's function matrix with the matrix element vector and its hermitian conjugate. In this case, we expect that the TR-ARPES signal will be the most physical result and will maintain nonnegativity. If it becomes negative, this would then be a fault of the constant matrix element approximation which would need to be fixed to correct the error.

The constant matrix element approximation is much more robust for the case of angle-integrated spectra. In this situation, 
the final formula in Eq.~(\ref{eq: final_pes}) holds, one can assume the matrix elements are the same for each band, and
if the local Green's function is diagonal in the band basis, then the signal is proportional to the trace of the local lesser Green's function.

\section{Discussion}

In this work, we helped clarify a number of subtle points associated with TR-ARPES studies when the number  
of low-energy bands is larger than one. These effects become particularly important for materials with a basis like graphene, or most topological insulators. Our focus was on how one generalizes the constant matrix element approximation from
one to many bands. It turns out that while the single-band case is completely determined with no extra freedom, the case
with multiple bands is more complex and has extra degrees of freedom associated with them.

In particular, there can be a different matrix element associated with each band. But more than that, if the basis for the
Green's function is changed, then one is forced into considering matrix elements that are no longer constant, since the new basis relates to the old by a unitary transformation which may change with time. We also discussed that in the case when the Green's function is diagonal, then the constant matrix element approximation is proportional to the trace of the Green's function if all matrix elements are chosen to be the same. Finally, we described how one can make a further uncontrolled approximation of using the trace formula with the same constant matrix elements for different bases, but in that case, it is likely that the spectra will become negative for some times and frequencies.

We ended by discussing the case for the constant matrix element approximation when one considers angle-integrated spectra. Here there are far fewer degrees of freedom in the results, and they also are manifestly gauge-invariant.

In the future, it would be interesting to discuss both the constant matrix element approximation and the issue of gauge invariance within a fully developed theory starting from first principles rather than in an {\it ad hoc} fashion, as we currently do. It also would be useful and interesting to understand what functional dependence of the matrix elements is most critical for understanding real spectra. Is it the wavevector dependence or the band index dependence? How do these results affect
the final TR-ARPES spectra?

\section{Materials and Methods}

No materials were employed in this study. The methods were standard many-body physics calculations.

\vspace{6pt} 


\acknowledgments{
We thank useful discussions with Michael Sentef, Benjamin Fregoso, and Michael Kolodrubetz.
This work was supported
by the Department of Energy, Office of Basic Energy Sciences,
Division of Materials Sciences and Engineering
(DMSE) under Contract No. DE-FG02-08ER46542.
HRK acknowledges support from the DST, India, and JKF also acknowledges support by the McDevitt bequest at Georgetown.
}

\authorcontributions{J.K.F. conceived of the idea for this study and both authors wrote the manuscript.}

\conflictofinterests{The authors declare no conflict of interest. The funding sponsors had no role in the design of the study; in the collection, analyses, or interpretation of data; in the writing of the manuscript, and in the decision to publish the results'.} 

\abbreviations{The following abbreviations are used in this manuscript:\\

\noindent 
\begin{tabular}{@{}ll}
TR-ARPES & time-resolved angle-resolved photoemission spectroscopy\\
TR-PES & time-resolved photoemission spectroscopy\\
LEED & Low-energy electron diffraction\\
TRL & time-reversed LEED
\end{tabular}}


\bibliographystyle{mdpi}

\renewcommand\bibname{References}



\end{document}